**Observation of the Berezinskii-Kosterlitz-Thouless transition in Boron-doped diamond films**

Christopher Coleman, and Somnath Bhattacharyya*

Nano-scale Transport Physics Laboratory and DST/NRF Centre of Excellence in Strong Materials, University of the Witwatersrand, Private Bag 3, WITS 2050, Johannesburg, South Africa

The occurrence of the Berezinskii-Kosterlitz-Thouless (BKT) transition is investigated in heavily boron-doped nanocrystalline diamond films through a combination of current-voltage and resistance measurements. We observe a robust BKT transition in the nanocrystalline diamond films with smaller grain size along with transport features related to vortex pinning. The vortex core energy determined through analysis of the resistance temperature curves was found to be anti-correlated to the BKT transition temperatures. It is also observed that the higher BKT temperature is related to an increased vortex-antivortex binding energy derived from the activated transport regions. Further, the magnetic field induced superconductor insulator transition shows the possibility of the charge glass state. The consequences of granularity such as localization and vortex pinning can lead to tuneable BKT temperatures and strongly affects the field induced insulating state.

Superconductivity in heavily boron-doped nanocrystalline diamond (BNCD) films was discovered in 2004[1] and has in the last decade been intensely investigated due to some outstanding features such as a non- BCS tuneable critical point[2,3], zero bias anomalies[4] and possible Kondo effects[4]. Although many reports have been made on the possible pairing mechanism[5,6,7] the exact nature of electron correlations in the BNCD films is still actively being investigated. There is a growing amount of evidence both experimental and theoretical that suggest the diamond crystal grains are separated by a $sp^2$ hybridised carbon phase similar to disordered graphene[8,9,10] highlighting the possibility of Dirac fermion or chiral superconductivity in this system. Additionally the effect of structurally arranged boron in these graphitic boundaries has recently been researched with respect to inducing interfacial superconductivity[11]. Although the BKT transition has been reported in carbon nanotubes[12] and predicted for graphene[13], the vortex structure of BNCD films have been studied through scanning tunnelling microscopy[14] and yet to date no report on the BKT transition has been made for this material.

In this work we study a batch of BNCD films having a variation of grains size (details of the samples are presented in Table 1 and Ref. 4). We investigate the BKT transition (vortex-antivortex binding) in certain films and relate this temperature as well as the absence of transition in other certain films to pinning effects verified through resistance vs temperature (*R-T)* and current-voltage (*I-V)* analysis. Additionally we investigate the magnetic field induced superconductor-insulator transition where the density of field induced localized vortices increases up to a quantum critical point above which they condense into an insulating Bose glass state. This transition is expected for disordered superconductors[15], and is related to the charge vortex duality[16] which has been identified in other granular materials through a universal scaling analysis[16] which expects the collapse of all magnetoresistance isotherms to a single curve described by a critical exponent product ($\upsilon z$) of order unity. The observed features are again related to temperature dependent pinning or localization effects arising from the grain boundaries.

*I-V* characteristics at different temperature intervals for certain samples (B5, and to a lesser extent B4) show strong asymmetric features with respect to sweeping direction. Upon closer examination these hysteretic features in the *I-V* sweeps are found to be strongly temperature dependent as shown in Fig. 1 (a & inset), in fact the difference in forward and reverse critical current is found to persist up to approximately 1.3 K, above which there is no observable difference in forward and reverse scans. Hysteresis in the *I-V* sweeps such as these are generally attributed to pinning, this essentially is due to a build-up of voltage as a result of vortex motion in the un-pinned state. Fig. 1(b) shows the current voltage data as a *log-log* scaled plot at different temperatures, the slopes of the respective curves are used to determine the α parameter (where $V \propto I^{\alpha}$) for the different boron-doped diamond samples. According to the Kosterlitz-Thouless criterion the BKT transition temperature can be evaluated from the plot by determining the temperature at which the parameter α jumps from a value of 1 to 3. This is shown in figure 1(b inset) where samples B5 and B4 have transition temperatures of $T_{\mathrm{BKT}}$ = 1.31 K and 0.59 K respectively and sample B2.5 does not show this jump in our experimentally accessible temperature range. Additionally it can be noted that the smooth transition observed in α occurs over a range of approximately 0.6 K instead of the sudden sharp jump, this has been observed before in 2D superconductors where finite size effects are present[17].

Fig. 2(a) shows a comparison of the phase transition from normal to superconducting state for the three samples plotted as resistance vs. temperature. The solid red curve is a fit using the following interpolation formula[17,18,19] adapted from the Halperin-Nelsons equation for the paraconductivity region between $T_{\mathrm{BKT}}$ and the mean field temperature $T_{\mathrm{C}}$.

$$\frac{R_N}{R} = 1 + \left(\frac{2}{A} sinh \frac{b}{\sqrt{t}}\right)^2, \qquad [1]$$

Where $A$ and $b$ are fitting parameters and $t = (T - T_{BKT})/T_{BKT}$ . Here we have used the BKT transition temperatures obtained from the *I-V log-log* plots (Figure 1(b inset)) leaving only the values of $b$ and $A$ as free parameters determined by the fitting.

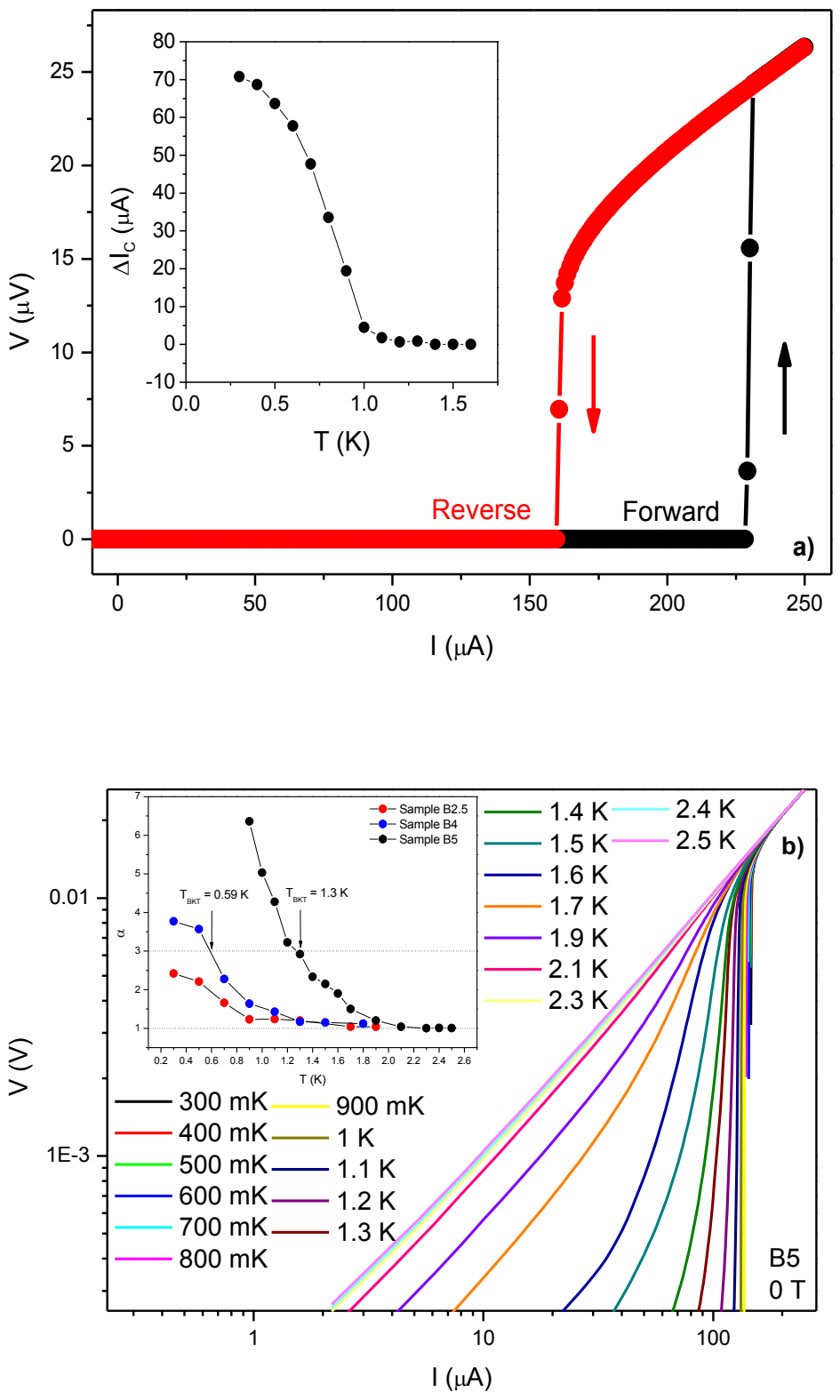

*Fig 1. (a) Current voltage characteristics of sample B5, there is a clear hysteresis feature due to depining of vortices, inset shows the magnitude of the difference in in the critical current for forward and reverse sweeps, the effects is diminished above 1K and is no longer observed above 1.3 K. (b) The log-log plot of Current Voltage sweeps, the slope of the curves are used to determine the KT critical exponent which is plotted as a function of temperature (inset) in order to determine the BKT transition temperatures.*

As the BKT transition was not observed for the B2.5 sample an estimated temperature of 0.1 K was used for comparative purposes. The interpolation formula fits the data relatively well except for two regions i.e. the flattening at the normal state and at the foot of the curve before the superconducting regime is fully reached. Both of these discrepancies have before been observed when fitting the Halperin-Nelson formula and are related to possible formation of a pseudo-gap[17] above the mean field critical point and finite size effects or inhomogeneity[18], respectively. It should be noted that although all three samples show to the same extent deviation from interpolation above the mean field temperature, sample B5 appears to be more adversely affected in the finite size effect regime which is expected to be due to smaller grain size. From the fit we extract the *b* parameter, this value is related to the vortex core energy and distance between the $T_{BKT}$ and $T_C$ points. For samples B5, B4 and B2.5 we extract *b*=2.85, 6.6 and 18.1, respectively.

*Table 1. Comparison of the properties of the three films used in this study including synthesis conditions, grain size, critical temperatures and coherence lengths.*

| Sample | $CH_4/H_2$ | Grain size | $T_C$ | $T_{BKT}$ | $\xi$ |
|---|---|---|---|---|---|
| B5 | 5 % | 30 nm | 1.79 K | 1.31 K | 10.8 nm |
| B4 | 4 % | 50 nm | 1.21 K | 0.59 K | - |
| B2.5 | 2.5 % | 70 nm | 0.98 K | - | 27.5 nm |

The values obtained for B5 and B4 are within the range of what has been reported for high temperature cuprates[19], however the value shown by B2.5 is unphysically large, in fact this is likely due to the hypothetical estimated BKT transition temperature used for the fitting. The *b* value can be related to the vortex core energy here expressed with respect to the value predicted by the XY model[17,18]. We find $\mu/\mu_{XY}$ = 2.37 and 3.27 for samples B5 and B4 respectively, thus the sample showing higher BKT transition temperature has a lower vortex core energy. This can again be related to the microstructure of the films if we take into account that vortex pinning, as expected for granular samples, can reduce the vortex core energy by an amount related to the pinning energy. As sample B5 has a finer grain structure, the vortex core energy is lowered and vortices nucleate more easily and are pinned at the grain boundaries.

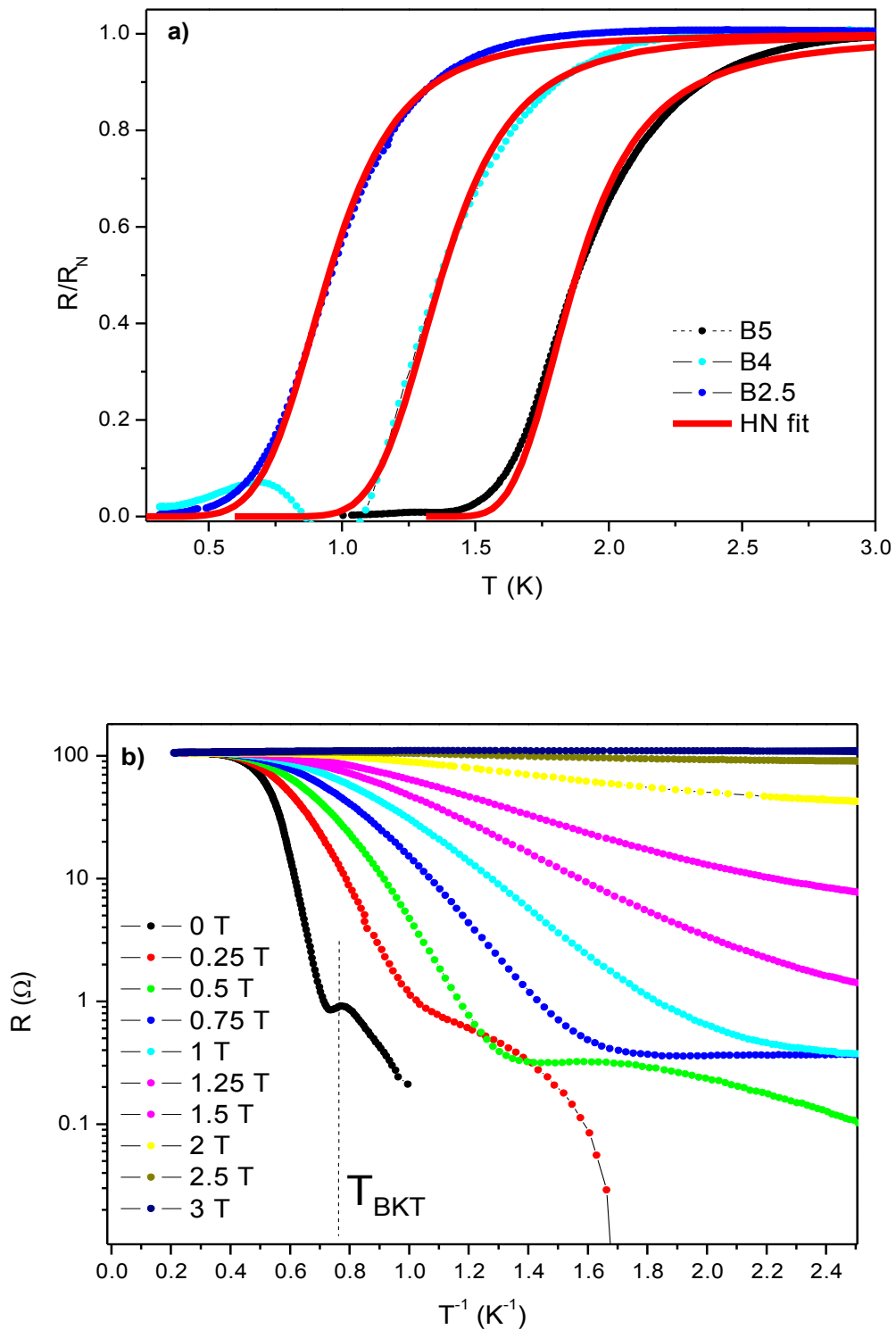

*Fig. 2 (a) the temperature dependent resistance for samples B5, B4 and B2.5. The red lines are a fit to the interpolation formula. Deviation from the fitting is most notable at the foot of the curves and related to finite size effects. (b) Resistance vs. inverse temperature characterising the Arrhenius activated region followed by a temperature independent regime at lower temperatures.*

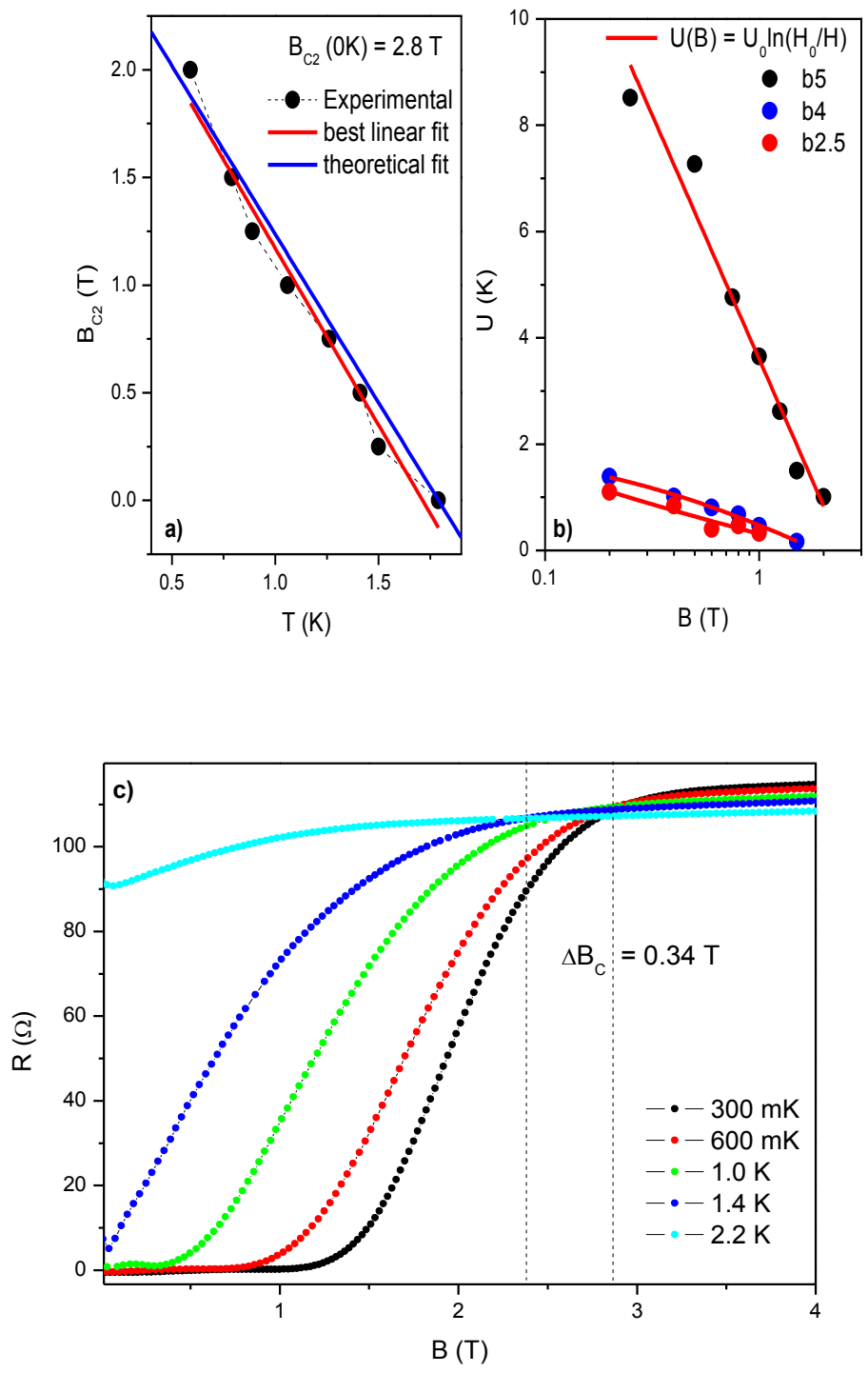

*Fig. 3 (a) The upper critical field derived from the RT data. The red line is the best fit used to obtain the Upper critical field B(0) = 2.8 T, the blue line is the theoretically expected behaviour. (b) The pairing energy obtained from R vs.$T^{-1}$. data, here the critical field was found to be 2.46 T. (c) Magnetoresistance data for sample B5, a smeared crossing point is observed between the two extracted critical fields*

We then examine the resistance as a function of inverse temperature as shown in figure 2(b). Here the black curve corresponds to the zero applied field measurement, it can clearly be seen that as temperature is lowered the curve exhibits a linear activated region which is interrupted by an up-turn or kink and then decreases further with lower temperatures. The peak of this kink is found to correspond exactly to the BKT transition temperature derived from the *I-V* data, this feature is thus identified as a signature of the bound vortex-antivortex transition. The BKT kink is however not observed in any instance when field is applied, signalling a field induced suppression of the transition, this can be related to an increase in field induced vortex density which is known to reduce the pairing energy.

We observe a trend in the data indicating a change from activated behaviour to some other phase identified by the drastic decrease in slope and eventual temperature independent state. This feature has been explained in light of Macroscopic Qunatum Tunneling (MQTV) of vortices[20] and more recently as a Quantum or Bose metal phase[21,22]. In a 2D superconductor the dissipation is expected to arise from vortex-antivortex dissociation and thus by fitting the activated region we can extract the vortex binding energy. The dependence of this pairing energy as a function of applied field is shown in Fig. 3(b). Here it can be seen that sample B5 has the highest pairing energy when compared to samples B4 and B2.5. From the field dependent pairing energy we can also extract $H_0$ ~ $B_{C2}$ found to be 2.46 T for sample B5. For comparison the $T_C$ field dependence can also be used to determine the upper critical ($B_{C2}$) this is shown in Fig. 3(a) where $B_{C2}$ (0 K) = 2.8 T. This critical field is approximately 12% larger than the value obtained from the activated transport fitting, this discrepancy (0.34 T) although relatively small indicate that the system does not strictly follow BCS theory. The blue line shows a fit to $B_{C2} = \varphi_0/(2\pi\xi(0)^2)(1-T/T_C)$ from which we extract a zero field coherence length of 10.84 nm (approximately half the grain size). As suggested previously there is a strong possibility of a field induced quantum phase, this is investigated through magnetoresistance measurement. A smeared crossing point can be seen in both samples B5 and B2.5 as indicated by Fig. 3(c). However upon closer examination of the *log-log* plot of the magnetoresistance (Fig. 4(a)) we observe that the crossing point of the various isotherms are concentrated at a single point at low temperature then start to drift only at higher temperatures most notable above 1.3 K for sample B5.

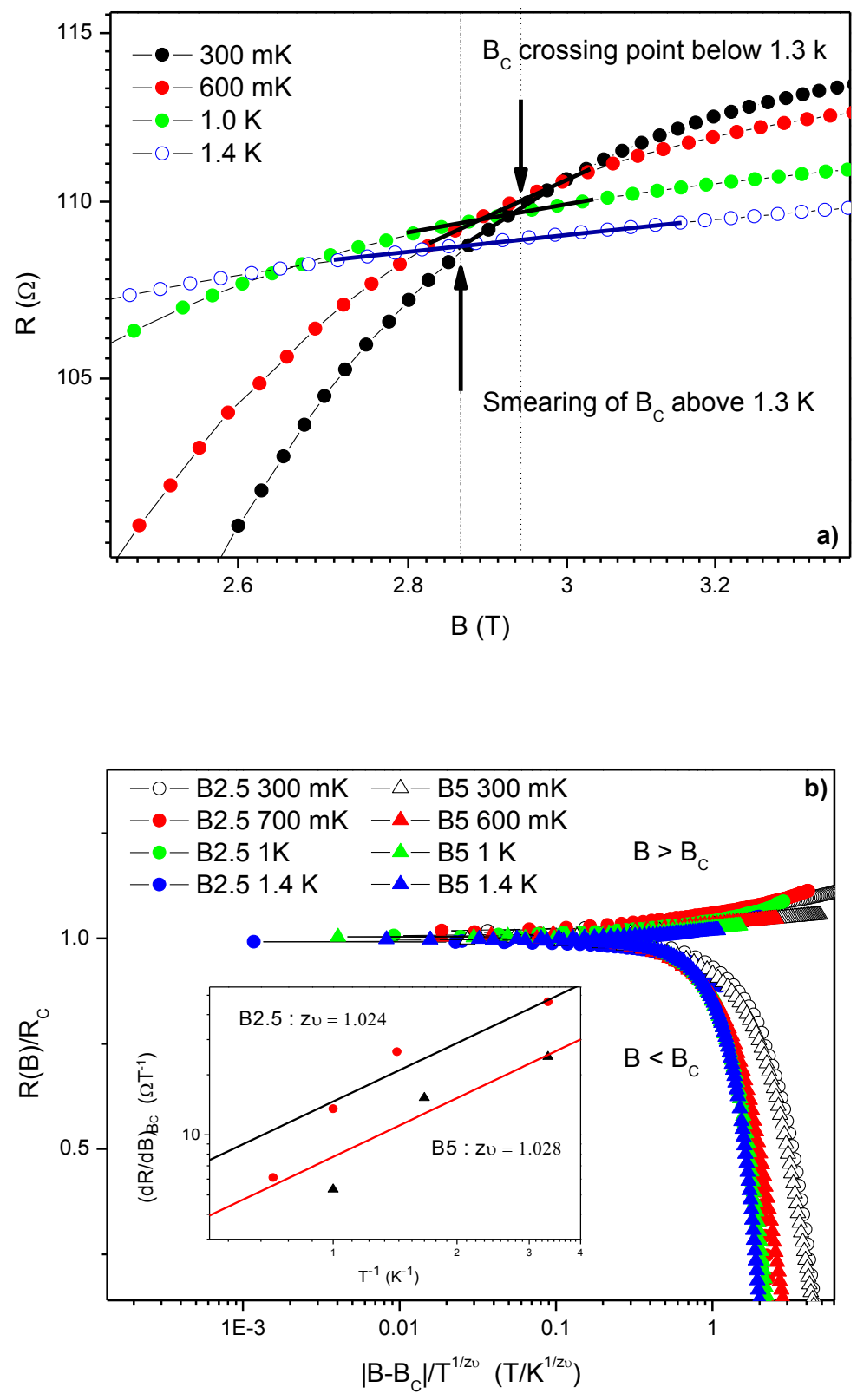

*Fig. 4 (a) The log-log plot of the magnetoresistance, notable features are the power law behaviour near the crossing point as well as deviation from crossing point for temperatures above 1.3 K. (b) The scaling analysis for samples B5 and B2.5, all data converge relatively well to single curve for critical exponents of zυ = 1.024 ± 0.3 and zυ = 1.028 ± 0.2 for samples B5 and B2.5 respectively, these values are confirmed by the best fit of slope of MR vs $T^{-1}$ as shown in the inset.*

This temperature was determined to be a significant point from both *I-V* and resistance temperature measurements. This is a significant observation as magnetoresistance isotherms having a single crossing point are widely studied in relation to the charge-vortex duality[14]. As seen in Fig. 4(b), the resistance is plotted using the universal scaling relation as usually demonstrated for the H-SIT. Both samples B5 and B2.5 collapse reasonably well onto a single curve, there are however slight deviations which are notable. This includes the tendency for lowest temperature lines of both samples to deviate from the collapse, this is indicated by the open circles and triangles in Fig. 4(b) compared to all other isotherms which collapse reasonably well to the same line. This may signify a crossover to alternative scaling behaviour at lower temperatures. It can also be observed that the insulating side of the scaling does not follow the steep upturn usually observed for SIT related superconductors and the value of the critical resistance is well below the theoretically predicted value[15,24]. Similar scaling behaviour has been observed before and was related to normal state electrons that contribute significantly to the transport through tunnelling[23], essentially the localization of charge carriers within the grain is reduced as temperature increases and thus the insulating state is not as severe. It has also been shown that the breakdown of the universal scaling can occur due to finite size effects and the formation of Josephson networks at certain grain size[24]. Nevertheless the scaling behaviour here results in critical exponent products of $z\upsilon = 1.024 \pm 0.3$ and $z\upsilon = 1.028 \pm 0.2$ for samples B5 and B2.5 respectively. These values are well within the theoretical prediction of unity[13] for a two dimensional system and are confirmed by the best fit to the $dR/dB$ isotherm data taken at the critical field $B_C$ (figure 4 (b) inset). These observations are a good indication that this system may very well exhibit the charge glass state above the H-SIT as predicted for disordered 2D superconductors[15].

In conclusion we have investigated the quantum transport of boron doped diamond and verified the BKT transition through *I-V* analysis and also identified features of this transition in the RT data. We find that the occurrence of the BKT transition is related to the enhanced pinning of the films with smaller grain size. We further investigate the field induced superconductor-insulator transition and find features of the charge glass state. The magnetoresistance crossing point is found to smear out above 1.3 K which was identified as the BKT temperature in zero magnetic field. This feature was also related to the reduced pinning, this time of the condensed Cooper pairs (charge glass state) or normal state electrons in the insulating phase of the H-SIT. Our findings thus highlight the importance of localization of vortices for the induction of the BKT transition as well as for Cooper pairs and normal state electrons in the insulating side of the H-SIT. This work shows the possibility of realizing topological superconductivity in carbon system which may be applied in quantum computing technology.

SB is very thankful to Prof. M. Nesledek for providing the samples and the NRF/NFP for the financial support.